\input harvmac

\overfullrule=0pt
\def\Title#1#2{\rightline{#1}\ifx\answ\bigans\nopagenumbers\pageno0\vskip1in
\else\pageno1\vskip.8in\fi \centerline{\titlefont #2}\vskip .5in}

\font\ticp=cmcsc10

\font\secfont=cmcsc10

%
%defs
%
\baselineskip=18pt plus 2pt minus 2pt

\def\ajou#1&#2(#3){\ \sl#1\bf#2\rm(19#3)}
%
%Caligraphic
\def\CH{{\cal H}}

%
%\def\eightbig#1{{\hbox{$\textfont0=\ninerm\textfont2=\niesy\left#1\vbox
% to6.5pt{}
%\right.\n@space$}}}
%\eightbig{N}

%Roman

%

%Greek

\def\a{\alpha}
\def\b{\beta}

\def\p{\pi}

%

%This paper

\def\TrH#1{ {\raise -.5em
                      \hbox{$\buildrel {\textstyle  {\rm Tr } }\over
{\scriptscriptstyle \CH _ {#1}}$}~}}

\def\IZ{\relax\ifmmode\mathchoice
{\hbox{\cmss Z\kern-.4em Z}}{\hbox{\cmss Z\kern-.4em Z}}
{\lower.9pt\hbox{\cmsss Z\kern-.4em Z}}
{\lower1.2pt\hbox{\cmsss Z\kern-.4em Z}}\else{\cmss Z\kern-.4em Z}\fi}
\def\IC{\relax\hbox{$\inbar\kern-.3em{\rm C}$}}
\def\IR{\relax{\rm I\kern-.18em R}}
\def\1{\relax 1 { \rm \kern-.35em I}}
\font\cmss=cmss10 \font\cmsss=cmss10 at 7pt

%

%Shortforms
\def\frac#1#2{{#1 \over #2}}

\def\p+{{\partial_+}}

%

%refs

%

\Title{\vbox{\baselineskip12pt
\hbox{\ticp CALT-68-2085}
\hbox{hep-th/9611119}
}}
{\vbox{\centerline {\bf ORIENTIFOLD AND F-THEORY DUALS OF CHL STRINGS} }}

\centerline{{\ticp
Jaemo Park\footnote{$^1$}{e-mail: jpk@theory.caltech.edu,
jaemo@cco.caltech.edu}
}}

\vskip.1in
\centerline{\it Lauritsen Laboratory of  High Energy Physics}
\centerline{\it California Institute of Technology}
\centerline{\it Pasadena, CA 91125, USA}

\vskip .1in

\bigskip
\centerline{ABSTRACT}
\medskip

 We present orientifold and F-theory duals of the heterotic
string compactification constructed by Chaudhuri, Hockney and
Lykken (CHL) which has the maximal supersymmetry but gauge
group of reduced rank. The 8-dimensional dual is given by
the Type IIA orientifold on the M\"{o}bius band.
We show the non-trivial monodromy on the base induces
non-simply laced gauge groups on the F-theory orbifolds
dual to the CHL strings.
The F-theory models dual to the CHL strings  in six dimensions
are examples of N=2 F-theory
vacua. We discuss the other N=2 F-theory vacua in six dimensions.

\bigskip

\bigskip
%\baselineskip=20pt plus 2pt minus 2pt
%\draft
\Date{November, 1996}

\vfill\eject

%References
\def\npb#1#2#3{{\sl Nucl. Phys.} {\bf B#1} (#2) #3}
\def\plb#1#2#3{{\sl Phys. Lett.} {\bf B#1} (#2) #3}
\def\prl#1#2#3{{\sl Phys. Rev. Lett. }{\bf #1} (#2) #3}
\def\prd#1#2#3{{\sl Phys. Rev. }{\bf D#1} (#2) #3}

\def\mpl#1#2#3{{\sl Mod. Phys. Lett. }{\bf #1} (#2) #3}

%-------------------
% references
%-------------------
%
\lref\Orie{
A. Sagnotti, in Cargese '87, ``Non-perturbative Quantum
Field Theory,'' ed. G. Mack et. al., (Pergamon Press, 1988)
p. 521\semi
P. Horava, \npb{327}{1989}{461}\semi
J. Dai, R. G. Leigh, and J. Polchinski,
\mpl{A4}{1989}{2073}.}

\lref\Sagn{M. Bianchi and A. Sagnotti,
\plb{247}{1990}{517}; \npb{361}{1991}{519}}

\lref\CHL{S. Chaudhuri, G. Hockney and  J. Lykken,
``Maximally Supersymmetric String Theories in D $<$ 10,''
\prl{75}{1995}{2264}}

\lref\ChPo{S. Chaudhuri and J. Polchinski,
``Moduli Space of CHL Strings,'' \prd{52}{1995}{7168}}

\lref\ScSe{J. H. Schwarz and A. Sen, ``The Type IIA Dual of the
Six-Dimensional CHL Compactification,'' \plb{357}{1995}{323},
  hep-th/9507027}

\lref\ChLo{S. Chaudhuri and D. Lowe,
``Type IIA-Heterotic Duals with maximal Supersymmetry,''
 \npb{459}{1996}{113}, hep-th/9508144}

\lref\ChLoII{S. Chaudhuri and D. Lowe,
``Monstrous string-string duality,''
\npb{469}{1996}{21}, hep-th/9512226}

\lref\DaPaI{
A. Dabholkar and J. Park, ``An Orientifold of Type IIB theory on K3,''
\npb{472}{1996}{207}, hep-th/9602030.}

\lref\DaPaII{
A. Dabholkar and J. Park, ``Strings on Orientifolds,''
\npb{477}{1996}{701},
hep-th/9604178.}

\lref\DaPaIII{
A. Dabholkar and J. Park, ``A Note on Orientifold and F-theory,''
\plb{394}{1997}{302},
hep-th/9607041.}

\lref\GiJoI{
E. Gimon and C. Johnson, ``K3 Orientifolds,'' \npb{477}{1996}{715},
hep-th/9604129.}

\lref\GiJoII{
E. Gimon and C. Johnson, ``Multiple Realizations of N=1 Vacua in
Six-Dimensions,'' \npb{479}{1996}{285}, hep-th/9606176.}

\lref\SenIII{A. Sen, ``F Theory and Orientifolds,''
\npb{475}{1996}{562},
hep-th/9605150.}

\lref\Aspi{P. Aspinwall, ``Some Relationships Between
Dualities in String Theory,'' CLNS-95-1359, hep-th/9508154}

\lref\VafaI{C. Vafa, ``Evidence for F Theory,''
HUTP-96-A004, \npb{469}{1996}{403}, hep-th/9602022.}

\lref\VaWi{C. Vafa and E. Witten,
 ``Dual String Pairs with N=1 and N=2 Supersymmetry in
Four Dimensions,'' hep-th/9507050}

\lref\MoVaI{D. Morrison and C. Vafa,
``Compactifications of F-Theory on
Calabi-Yau Threefolds-I,''  \npb{473}{1996}{74},
hep-th/9602114.}

\lref\MoVaII{D. Morrison and C. Vafa,
``Compactifications of F-Theory on
Calabi-Yau Threefolds-II,'' \npb{476}{1996}{437},
 hep-th/9603161.}

\lref\PolcI{J. Polchinski, \prl{75}{1995}{4724}.}

\lref\PolcII{J. Polchinski, ``Recent Results in String Duality,''
NSF-ITP-95-157, hep-th/9511157.}

\lref\PCJ{J. Polchinski, S. Chaudhuri and C. Johnson,
``Notes on D-Branes,''
 NSF-ITP-96-003, hep-th/9602052.}

\lref\PoWi{J. Polchinski and E. Witten,
``Evidence For Heterotic-type I String Duality,''
\npb{460}{1996}{525}, hep-th/9510169.}

\lref\GiPo{E.~G.~Gimon and J.~Polchinski, ``Consistency Conditions
for Orientifolds and D-manifolds,'' \prd{54}{1996}{1667},
 hep-th/9601038.}

\lref\Mo{This possiblity has been considered in informal discussions
but we are not certain of its origin.}

\lref\WittII{E.~Witten, ``Small Instantons in String Theory,''
\npb{460}{1996}{541}.}

\lref\HoWi{P. Horava and E. Witten, ``Heterotic and Type I String
Dynamics from Eleven-Dimensions,'' \npb{460}{1996}{506}, hep-th/9510209.}

\lref\DaMu{K. Dasgupta and S. Mukhi, ``F Theory at Constant
Coupling,'' \plb{385}{1996}{125}, hep-th/9606044.}

\lref\GaMo{O. Ganor, D. Morrison and N. Seiberg, ``Branes,
Calabi-Yau Spaces, and Toroidal compactification of the N=1
Six-dimensional $E_8$ theory,''  PUPT-1646, hep-th/9610251.}

\lref\GrHa{P. Griffiths and J. Harris, {\it Principles of Algebraic
Geometry}, Wiley-Interscience, New York 1978, p585.}

\lref\Beet{M. Bershadsky, K. Intriligator, S. Kachru, D.R. Morrison,
V. Sadov and C. Vafa, ``Geometric Singularities and Enhanced Gauge
Symmetries,''  HUTP-96-A107, hep-th/9605200.}

\lref\AspiII{P. Aspinwall and M. Gross , ``The SO(32)
Heterotic String on a K3
Surface,'' CLNS-96-1409, \plb{387}{1996}{735}, hep-th/9605131.}

Along with the recent development of string duality,
new ways of constructing string vacua have appeared.
These new constructions make manifest  some nonperturbative
aspects of the string theory which the previously known
constructions cannot see. Typical examples are orientifolds
and F-theory.

An orientifold is a generalization of orbifolds in which
the orbifold symmetry includes the orientation reversal
on the world sheet\refs{\Orie, \PolcI, \GiPo, \PCJ}.
Type I theory is an example of an orientifold
where the only symmetry gauged is the orientation reversal
of the Type IIB theory. An important application of orientifolds
is the construction of models in 6-dimensions with
N=1 supersymmetry \refs{\Sagn, \DaPaI, \GiJoI, \DaPaII,
\DaPaIII}. The models with multiple tensor multiplets
can be easily constructed using the orientifold, while
this is not possible in the conventional Calabi-Yau
compactification which gives only one tensor multiplet.
Similarly, small instantons \WittII, which cannot be described as
a conformal field theory in heterotic string theory, have
a perturbative description in terms of Dirichlet 5-branes
in the dual orientifold.

On the other hand, F-theory is a new way of compactifying
Type IIB theory in which the complex coupling $\lambda$ of
Type IIB theory is allowed to vary over the compactified space
 \refs{\VafaI, \MoVaI, \MoVaII}. The complex coupling can be seen
as the complex structure parameter of the elliptic fibration
over the base $B$ on which the Type IIB theory is compactified.
The coupling can undergo non-trivial $SL(2,Z)$ transformations
as we move along non-trivial cycles on the base $B$.
Since the nonperturbative $SL(2,Z)$ symmetry of the Type IIB
theory is realized as
the $SL(2,Z)$ transformation of the elliptic fibration, F-theory is quite
powerful in studying nonperturbative phenomena
in string theory such as the phase transition involving
tensionless strings in 6-dimensions.

In another interesting development, Chaudhuri, Hockney and
Lykken (CHL) have constructed new examples of the heterotic
string compactification with maximal supersymmetry but
with gauge groups of reduced rank \refs{\CHL, \ChPo}. It turns out that
all CHL models can be identified with toroidal
compactification of the heterotic string theory further
modded by some discrete global symmetry such as the
interchange of the two $E_8$'s of the heterotic string
theory \PolcII. Some duality aspects of CHL models are known.
The M-theory and Type IIA duals of various CHL models
in 6-dimensions and below
were investigated in detail \refs{\ScSe, \ChLo, \Aspi}.

The purpose of this letter is to find the dual theory
of a CHL model to be described later, using the orientifold and F-theory.
Type IIA orientifold on the M\"{o}bius band is dual to the
CHL model in 8-dimensions.
In the strong coupling limit, this configuration is lifted
to M-theory compactified on the M\"{o}bius band which gives
the 9-dimensional dual of the CHL string. We present the F-theory
orbifold dual with non-simply laced gauge groups.
The F-theory dual of the CHL model in 6-dimensions
constitutes a part of the F-theory vacua in 6-dimensions with N=2
supersymmetry hitherto uninvestigated. We discuss the other
N=2 F-theory vacua in 6-dimensions and give the
orientifold duals in simple cases.

The initial CHL models are given by the free fermionic construction\CHL,
\ but Chaudhuri and Polchinski\ChPo \ have constructed one of these models
as an asymmetric orbifold. They considered $Z_2$ orbifold of the
toroidally compactified heterotic string theory, where $Z_2$ action
interchanges the two $E_8$ components of the momentum lattice,
together with a half shift on a compactified circle.
Under this $Z_2$ modding, only the symmetric combination of $E_8$'s
survives, thereby reducing the the rank of the gauge group by eight.
The $Z_2$ modding is possible if the two $E_8$'s are broken in an
identical manner. Similar construction can be done in $SO(32)$
heterotic string theory. Since $SO(32)$ and $E_8 \times
E_8$ heterotic string theory are equivalent upon compactification
on a circle, the $Z_2$ orbifolds on both sides are on the same moduli
space. Since the $Z_2$ symmetry adopted in the orbifold
construction is a freely acting $Z_2$, there are no massless states
in the twisted sector at generic points of the moduli space.
We are mainly interested in this $Z_2$ orbifold example.

Since there is a duality conjecture on Type I and heterotic string
theory \PoWi,
we expect that the dual theory of the CHL model can be found
in Type I side. The strategy we will take is the following:
Using the duality relation between  Type I and the heterotic
theory, we can figure out the $Z_2$ symmetry in Type I which corresponds
to the $Z_2$ symmetry used in the CHL construction in the
heterotic theory.
If the $Z_2$ action
used is freely acting,  the adiabatic argument\VaWi\
assures us  of the duality
between the $Z_2$ orbifolds.

It is more convenient to work with Type ${\rm I}^{\prime}$ theory,
because
the $Z_2$ action is realized geometrically.
Since the gauge groups are realized as the Chan-Paton degrees of
freedom of 8-branes in the Type ${\rm I}^{\prime}$ theory,
$Z_2$ action should interchange the
8-branes and this must be
accompanied by a half shift along a circle which
guarantees the total action
freely acting. This means that the total
$Z_2$ action does not produce
additional orientifold planes which contribute to non-zero tadpole,
as we will see shortly.
We construct the 8-dimensional orientifold.
Consider Type IIA theory compactified on the torus,
say in the $X^8$ and
$X^9$ directions with the identification
$X^8 \equiv X^8 + 2 \pi r_8$
and  $X^9 \equiv X^9 + 2 \pi r_9$.
We take an orientifold with the projection
\eqn\orbi{
\quad  \{1, \eta_8\eta_9, R_9\Omega,
\eta_8\eta_9 R_9\Omega \},}
 where $\eta_i$ denotes a half shift along the
$i$-th circle; $X^i \rightarrow X^i + \pi r_i, i=8,9$.
This is the Type ${\rm I}^{\prime}$ theory modded
by the $Z_2$ action $\eta_8\eta_9$.
Note that $\eta_9$ relates an 8-brane located at $X_9=X_{9_0}$ to
an 8-brane at $X_9=X_{9_0} + \pi r_9$, and $\eta_8$ is the accompanying
shift. The action $\eta_8\eta_9$ preserves all of the harmonic forms
on the torus, hence the supersymmetry is not reduced.
This model has the same supersymmetry as Type
${\rm I}^{\prime}$ theory
toroidally compactified to 8-dimensions. The action of $\eta_8\eta_9$
on the oscillator modes is trivial. For the ground states
$|p_i, L^i \rangle$ without oscillations, which have the
quantized momenta $p_i\equiv m_i/R_i$ with $m_i$ integer valued
in the compact directions,
and winding $L^i\equiv X^i(\pi)-X^i(0)=2\pi w^i R_i$
with $w^i$ integer, $\eta_8\eta_9$
has the action
\eqn\action{\eta_8\eta_9 |p_i, L^i \rangle = (-1)^{m_8} (-1)^{m_9}
|p_i, L^i \rangle.}
The massless modes of the
closed string sector coincide with those of Type
${\rm I}^{\prime}$ theory
since $\eta_8\eta_9$ acts trivially on those modes.
The twisted sector of the closed string has half-integer winding
modes in $X^8, X^9$ directions with the momentum modes and the
oscillator modes unchanged.

We can determine the open string spectrum by calculating the
tadpoles \GiPo. The open string sector arises from the addition of the
D-branes to cancel the tadpole in the Klein bottle amplitude.
The Klein bottle amplitude consists of two parts, one from the
trace evaluation with $R_9\Omega$ and the other from the trace with
$\eta_8\eta_9 R_9\Omega$. The loop channel amplitude of the
former gives the tadpole which requires 32 8-branes for the
tadpole cancellation as in Type ${\rm I}^{\prime}$ theory.
The loop channel momentum sum of the latter
is proportional to $\Sigma_{m_8} (-1)^{m_8}
e^{\frac{-\pi t \alpha' m_8^2}{r_8^2}}$, where $t$ is the loop channel
parameter. This gives vanishing tadpole in the tree channel
in the $t\rightarrow 0$ limit, as one can see using the Poisson
resummation formula. The Klein bottle amplitude of the twisted
sector also vanishes, since the half winding modes in the $X^8$
direction are odd under $R_9\Omega$ and $\eta_8\eta_9 R_9\Omega$.
Thus we have 32 8-branes in all. Because of the orientifold
projection, only orthogonal gauge groups are allowed.
In addition, brane configuration should be
invariant under the action $\eta_8\eta_9$. The maximal
gauge group is obtained if we put the 16 8-branes at $X^9=X^9_0$
and the other 16 8-branes at $X^9=X^9_0+\pi r_9$. The gauge group
in this case is $SO(16)$. It is clear that we can obtain the
orthogonal subgroups of $SO(16)$ by locating 16 branes at different
positions and locating the other 16 branes compatible with
the action $\eta_8\eta_9$. From the initial Type
${\rm I}^{\prime}$ theory,
we obtain the model whose rank of the gauge group is reduced
by eight.
Thus we see that this orientifold construction gives the same
massless spectrum as the CHL model of the $SO(32)$ heterotic
string theory where interchange of the momentum lattice
is accompanied by the half-shift along the $X^8$ direction.

Note that the above orientifold action \orbi\  turns the compactified
torus into the M\"{o}bius band. Hence the CHL model in 8-dimensions
is dual to the Type-IIA orientifold compactified on the M\"{o}bius band.
By considering the strong coupling limit
we can lift this construction to M-theory compactified
on the M\"{o}bius band, which is dual to the CHL model
in 9-dimensions with the gauge group $E_8$.

It was explained in\PCJ\  how to obtain the $E_8\times E_8$ M-theory
of Horava and Witten \HoWi, starting from Type
${\rm I}^{\prime}$ theory.
We put 14 8-branes on each fixed point and place
the other 4 branes in symmetric fashion with respect to the fixed
points so that the resulting
gauge group is $(SO(14)\times U(1))^2$. The configuration
is dual to $SO(32)$ heterotic string theory compactified on
a circle with a particular Wilson line where the enhanced
gauge group $E_8\times E_8$ is
achieved at a particular radius of the compactified circle.
The dual configuration in the Type ${\rm I}^{\prime}$ side
is obtained
by taking the strong coupling limit.
In this limit, the other 4 branes approach the fixed points
and additional 0-brane states become massless to form
adjoint of $E_8\times E_8$. In this limit, the radius of
the compactified circle goes to infinity. The relation between
the Type ${\rm I}^{\prime}$
theory and the M-theory on $S^1/Z_2\times S^1$ is given by
\eqn\rel{
R_{I'}=r_1 r_2^{1/2}, \quad
g_{I'}=r_2^{3/2},}
where $R_{I'}$ and $g_{I'}$ are the compactified radius
and the coupling constant of Type ${\rm I}^{\prime}$ theory
respectively,
$r_1$ is the radius of $S^1/Z_2$ and $r_2$ is the radius of
$S^1$ of M-theory. The limit we take is
$r_2 \rightarrow \infty$ limit which corresponds
to $g_{I'},R_{I'} \rightarrow
\infty$ limit.

Since the strong coupling limit
is compatible with the action $X^9\rightarrow X^9+\pi r_9$,
we can  take the same limit for the above Type IIA orientifold.
In this limit, we obtain the $Z_2$ orbifold of $E_8\times E_8$
 M-theory where $E_8$
exchange is accompanied by the shift $\eta_8$. By changing
the coordinate label $X^8$ to $X^9$, we obtain the M-theory
compactification in 9-dimensions and the compactified space
is the M\"{o}bius band. This is dual to the CHL model in
9-dimensions with the gauge group $E_8$ where $E_8$ exchange
is accompanied by the shift on a circle. The duality can
be directly argued using the adiabatic argument and
the duality between the M-theory
on $S^1/Z_2$
and the $E_8\times E_8$ string theory\Mo.

Now we turn into the F-theory dual of the CHL model
in 6-dimensions.
Since there is a conjectured duality in 8-dimensions between
the F-theory on $K3$ and the heterotic string theory
on $T^2$ \refs{\VafaI,
\SenIII},
we expect that one can construct the F-theory dual by modding out
the $Z_2$ symmetry which corresponds to the $Z_2$ symmetry
of the CHL string.
Since the gauge group of the F-theory appears as singular elliptic
fibers\refs{\MoVaI,\MoVaII},
the $Z_2$ symmetry interchanging the gauge group is realized
as a $Z_2$ involution of K3 which interchanges the singular fibers.
This involution should respect the fiber structure.
We start with the particular K3 orbifold $T^4/Z_2$. F-theory on this
particular orbifold was considered by Sen in establishing the duality
between F-theory and the orientifold which is obtained by
T-dualizing the Type I theory along the $X^8$ and $X^9$
directions\SenIII. If the F-theory model
has the constant coupling as in the $T^4/Z_2$ orbifold,
the corresponding orientifold configuration should
satisfy the local tadpole cancellation so that the resulting
orientifold configuration has the constant coupling.
Let us denote the complex coordinates of the six-torus by
$z_1,z_2,z_3$ with identification $z_l\equiv z_l+1 \equiv
z_l+i, l=1,2,3.$ We consider the $Z_2 \times Z_2$ orbifold of
F-theory with the following generators $\alpha, \beta$
of the orbifold action.
\eqn\orbia{\eqalign{
\a \quad : & \quad (z_1, z_2, z_3) \rightarrow (-z_1, -z_2, z_3), \cr
\b \quad : & \quad (z_1, z_2, z_3) \rightarrow (-z_1, -z_2+\frac{1}{2},
 z_3+\frac{1}{2}). }}
Here $z_1$ denotes the coordinate of the elliptic fiber and
$2\pi r_2 z_2\equiv X^8+i X^9, 2\pi r_3 z_3 \equiv X^6+i X^7$.
Restricted
to $z_1,z_2$ coordinates, $\alpha$ is the $z_2$ action of $T^4/Z_2$
orbifold.
The torus parametrized by $z_2$ has four fixed points of $\alpha$.
The singular fiber over each fixed point is of $D_4$ type\foot{The
possible singularities of elliptic fibers are classified by
Kodaira. Those singularities fit into the ADE classification, which
in turn give ADE gauge groups in F-theory\refs{\MoVaI, \MoVaII.
}}, and
we have $SO(8)^4$ gauge group due to the four singular fibers.
The action $\beta$ is
a $Z_2$ involution on K3 orbifold combined with the shift along the
$z_3$ coordinate. Since the generator $\beta$ preserves the
holomorphic 2-form of K3 orbifold and the holomorphic form on the
third torus, this orbifold has the same supersymmetry as the F-theory
on $T^4/Z_2 \times T^2$, i.e., N=2 supersymmetry in 6-dimensions.
Four singular $D_4$ fibers are paired by the
action $\beta$ and the resulting gauge group is $SO(8)^2$.
Since $\beta$ is freely acting, there are no additional
massless modes. The massless spectrum consists of N=2 supergravity
multiplet and N=2 vector multiplet with gauge group $SO(8)^2$.

One can check that this conclusion  is consistent with the known duality
between the F-theory and the orientifold.
If we use the duality dictionary between the F-theory and the
orientifold\SenIII, $\alpha$ is mapped to $\Omega (-1)^{F_L} R_{89}$ and
$\beta$ is mapped to $\Omega (-1)^{F_L} R_{89}\eta_6\eta_8$.
Hence the orientifold projection corresponding to the above
orbifold is $\{ 1, \eta_6\eta_8, \Omega (-1)^{F_L} R_{89},
\Omega (-1)^{F_L} R_{89}\eta_6\eta_8 \}$.
By similar tadpole calculation as we did previously,
we can conclude that
this orientifold has the gauge group $SO(8)^2$ if the tadpole
cancels locally. If we T-dualize the orientifold
along the $X^9$ direction, the orientifold projection
becomes $\{1, \eta_6\eta_8, R_8\Omega, \eta_6\eta_8 R_8\Omega  \}.$
After some coordinate relabeling, we can see that this is
the original orientifold model of \orbi\
  compactified further on $T^2$.
Since this orientifold model is dual to the
CHL model, the above F-theory $Z_2 \times Z_2$ orbifold should be dual
as well.
Further evidence for the duality between the F-theory orbifold
and the CHL model can be seen if we compactify F-theory further
on a circle. This theory is on the same moduli space as M-theory
compactified on the above $Z_2 \times Z_2$ orbifold, according to
the duality between F-theory and M-theory\VafaI. This M-theory
orbifold is a special case of the M-theory on $(K3\times T^2)/Z_2$
which is dual to the CHL model in 5-dimensions as argued in \ScSe.
The $Z_2$
action is the half-shift along the torus combined with the involution
of $K3$ under which eight anti-self-dual
2-forms of K3 are odd and
the remaining harmonic forms are even. This $Z_2$ involution
is realized by $\beta$ in the $Z_2\times Z_2$ orbifold of \orbia.
Thus we see that the duality between the M-theory on $(K3\times T^2)/Z_2$
and the CHL model in 5-dimensions is lifted to the duality between
the F-theory on $(K3\times T^2)/Z_2$ and the CHL model in 6-dimensions.

One can consider
the models corresponding to other points of the moduli
space of F-theory on $(K3\times T^2)/Z_2$.
One such model is  given by the orbifold generated by the following
action.
\eqn\orbib{\eqalign{
\a \quad : & \quad (z_1, z_2, z_3)
\rightarrow (iz_1, -iz_2, z_3), \cr
\b \quad : & \quad (z_1, z_2, z_3)
\rightarrow (-z_1, -z_2+\frac{1+i}{2},
 z_3+\frac{1}{2}). }}
Restricted to $z_1, z_2$ coordinates, $\alpha$ is the $Z_4$ action
of the $T^4/Z_4$ orbifold. F-theory on $T^4/Z_4$ was considered by
Dasgupta and Mukhi\DaMu. The torus parametrized by $z_2$ has four
fixed points
under $\alpha^2$. Two of them are fixed points of $\alpha$ as well,
but the other two form a doublet under $\alpha$. The singular
fiber over each fixed point of $\alpha$ is of $E_7$ type. The other
singular fiber over the doublet under $\alpha$ is of $D_4$ type.
The F-theory on $T^4/Z_4$ has the gauge group $E_7 \times E_7
\times SO(8)$. The action $\beta$ is a $Z_2$ involution accompanied
by a half-shift along the $z_3$-torus. Since $\beta$ preserves the
holomorphic 2-form of the K3 orbifold and the holomorphic
1-form of the $z_3$ torus, this orbifold has N=2 supersymmetry
in 6-dimensions. The special feature of $\beta$ is that it induces
nontrivial monodromy on the $D_4$ fiber while it interchanges
two $E_7$ fibers.

The enhanced gauge group of the singular fiber which
suffers the nontrivial monodromy along a nontrivial cycle of
 the base manifold was considered in detail in \refs{\Beet, \AspiII}.
The enhanced gauge group is the monodromy invariant part of the
apparent local gauge group. The action of the monodromy on the
blown-up fiber can be translated into an action on the Dynkin
diagram of the simply-laced gauge group. The required group is
the subgroup invariant under this outer automorphism.
The gauge group coming from the $D_4$ fiber with the monodromy
induced by $\beta$ is $SO(7)$. The gauge group of the above
model is $E_7 \times SO(7)$.
As explained by Dasgupta and Mukhi, we cannot give the perturbative
orientifold description of the above F-theory orbifold since
the orbifold action of $T^4/Z_4$ corresponds to
 the nonperturbative symmetry
of Type IIB theory. But we already established the duality between
F-theory on $(K3\times T^2)/Z_2$ and the CHL model at a particular
point of the moduli space, hence the above model is necessarily dual to
the CHL model in six dimensions.

Another model in the same moduli space is given by the
orbifold generated by the following action when we consider
the six-torus based on the hexagonal lattice i.e.,
$z_l\equiv z_l+1\equiv e^{\frac{2\pi i}{3}}z_l, l=1,2,3.$
\eqn\orbid{\eqalign{
\a \quad : & \quad (z_1, z_2, z_3)
\rightarrow (e^{\frac{2\pi i}{3}}z_1,
e^{\frac{-2\pi i}{3}}z_2, z_3), \cr
\b \quad : & \quad (z_1, z_2, z_3)
\rightarrow (-z_1, -z_2,z_3+\frac{1}{2}). }}
Restricted to $z_1,z_2$ coordinates, $\alpha$ is the $Z_3$ action of the
$T^4/Z_3$ orbifold\DaMu. The torus parametrized by $z_2$ has three
fixed points of $\alpha$. Each singular fiber over the fixed point
is of $E_6$ type. F-theory on $T^4/Z_3$ orbifold has the gauge
group $E_6 \times E_6 \times E_6$. The action $\beta$ interchanges
two $E_6$ fibers and induces a nontrivial monodromy
which reduces $E_6$ to $F_4$. The resulting gauge group is
$E_6 \times F_4$.

In the three F-theory orbifolds considered so far, the base of
each model is $(P^1\times T^2)/Z_2$. But clearly one can
consider more general situations. If we consider F-theory
compactified on $(K3\times T^2)/G$, the base is
$(P^1\times T^2)/G$. In order to have N=2 supersymmetry in
6-dimensions, we should restrict $G$ to be a finite
automorphism of $K3\times T^2$ compatible with the elliptic
fibration which preserves holomorphic
forms of $K3$ and $T^2$.

 As explained in \GaMo, the sublattice of $\Gamma^{(18,2)}$ of the
cohomology lattice of $K3$ determines an elliptic $K3$, where
$\Gamma^{(p,q)}$ denotes a lattice of signature $(p,q)$.
The lattice $\Gamma^{(18,2)}$ is obtained from
the cohomology lattice of $K3$, $H^2(K3) \oplus H^0(K3) \oplus H^4(K3)
\simeq \Gamma^{(19,3)} \oplus \Gamma^{(1,1)}$ by
splitting off the classes
of the base and fiber of the fibration. Conversely, given an
even-self-dual lattice of signature (18,2),
one can find the corresponding elliptic $K3$ by the global Torelli
theorem. Combined with additional $\Gamma^{(2,2)}$ lattice
associated with $T^2$, this gives $\Gamma^{(18,2)} \oplus \Gamma^{(2,2)}$
which is isomorphic to $\Gamma^{(20,4)}$ and $G$ acts on $\Gamma^{(20,4)}$
as a lattice isomorphism. This condition is the same one as we encounter
in the general CHL compactification of heterotic string theory
in six dimensions. Thus we expect that F-theory on $(K3\times T^2)/G$
is dual to the CHL compactification of the heterotic string theory
on $T^4/\tilde{G}$ where $G$ and $\tilde{G}$ act on the same way
on $\Gamma^{(20,4)}$.

If $G$ is freely acting, $G$ must act on $T^2$ by translation,
which implies $G$ should be an abelian group with at most two
generators.
Again if we compactify further on a circle, this
theory is on the same moduli space as the M-theory on
$(K3 \times T^2)/G$. This is indeed dual to other CHL constructions
as investigated in \ChLo, where $G$ acts on $K3$ as an abelian
symplectic automorphism, {\it i.e.}, an abelian automorphism preserving the
holomorphic 2-form of $K3$.
But more general configurations are possible. Such configurations
in the Type IIA side are considered in \ChLoII.
If we compactify F-theory on $(K3\times T^2)/G$ further on $T^2$,
this model is on the same moduli space as the Type IIA theory
on $(K3\times T^2)/G$, which is dual to the heterotic theory
on $T^6/G$. On the heterotic side, general $G$ acts on $\Gamma^{(22,6)}$
as an lattice isomorphism with general shifts. The general shifts
on the heterotic side is mapped to the choice of Ramond-Ramond
fluxes localized at the fixed points of $K3$ under the heterotic-Type IIA
duality. Thus the general configuration in the Type IIA side is
the orbifold modded by symplectic automorphism of $K3$ with the
Ramond-Ramond field background, which is not a conventional
superconformal field theory background. General F-theory
compactification can be thought to be the decompactifying limit of the
corresponding Type IIA configuration when $G$ is compatible with
the elliptic structure, which implies that $G$ acts nontrivially only
on $\Gamma^{(20,4)}$ sublattice of $\Gamma^{(22,6)}$.

Since we have specified some of the N=2 F-theory vacua in
6-dimensions, one might wonder what else can appear as N=2 vacua
of F-theory. It is explained in \MoVaII that allowable
bases for N=2 F-theory vacua are $K3, (P^1\times T^2)/G,$ and
hyperelliptic surfaces. F-theory on the base K3 is the F-theory
compactified on $T^2\times K3$. The fiber structure should be
trivial in order to retain the Calabi-Yau condition. This is just
Type-IIB theory on K3.
The only remaining category is F-theory having a hyperelliptic
surface as base. A hyperelliptic surface is a complex torus
modulo a finite group $G$ acting freely.
We can give a simple example of the F-theory which has a
hyperelliptic surface as base.
Consider the orbifold with the following $Z_2$ symmetry.
\eqn\orbic{\eqalign{
\alpha \quad : & \quad (z_1, z_2, z_3)
\rightarrow (-z_1, -z_2, z_3+\frac{1}{2}). }}
The action $\alpha$ restricted to $z_2,z_3$ coordinates
produces the hyperelliptic surface where the freely acting
group is $Z_2$. The holomorphic 2-form $dz_1 \wedge dz_2$ and
1-form $dz_3$ survive under $\alpha$, hence F-theory on this orbifold
has $N=2$ supersymmetry in 6-dimensions. Since $\alpha$ is freely
acting, there are no singular elliptic fibers. Thus massless
spectrum of this model is just non-chiral N=2 supergravity
multiplet. We can find the orientifold dual of this model.
Using the duality dictionary, we can map this orbifold into
the orientifold with the group
$\{ 1, \Omega (-1)^{F_L} R_{89}\eta_6 \}$. After a similar
calculation as we did earlier, one can check that
the massless spectrum
of the orientifold agrees with that of the F-theory orbifold.
For other hyperelliptic surfaces\foot{There are seven
 distinct classes of hyperelliptic surfaces.
Those are explained in \GrHa.},
we can perform the similar analysis. Since hyperelliptic surfaces
are quotients of torus by freely-acting action,
there are no singular fibers for the associated elliptic
three-fold. The massless spectrum is again
N=2 supergravity multiplet in 6-dimensions.

\bigskip
\leftline{ \secfont Acknowledgements}
\bigskip

We would like to thank D.Lowe for helpful discussions
and D. Morrison
for valuable correspondence and discussions.
This work was supported in part by the U. S. Department of Energy
under Grant No. DE-FG03-92-ER40701.
\vfill
\eject
\listrefs
\end